# Pressure induced metallization and possible unconventional superconductivity in spin liquid NaYbSe$_2$

Zheng Zhang,[1,2,*] Yunyu Yin,[1,*] Xiaoli Ma,[1,*] Weiwei Liu,[1,2] Jianshu Li,[1,2] Feng Jin,[1] Jianting Ji,[1] Yimeng Wang,[1,2] Xiaoqun Wang,[3] Xiaohui Yu,[1,‡] Qingming Zhang[4,1,§]

[1] Beijing National Laboratory for Condensed Matter Physics, Institute of Physics, Chinese Academy of Sciences, Beijing 100190, China

[2] Department of Physics, Renmin University of China, Beijing 100872, China

[3] Department of Physics and Astronomy, Shanghai Jiao Tong University, Shanghai 200240, China

[4] School of Physical Science and Technology, Lanzhou University, Lanzhou 730000, China

## Abstract

Beyond the conventional electron pairing mediated by phonons, high-temperature superconductivity in cuprates is believed to stem from quantum spin liquid (QSL). The unconventional superconductivity by doping a spin liquid/Mott insulator, is a long-sought goal but a principal challenge in condensed matter physics because of the lack of an ideal QSL platform. Here we report the pressure induced metallization and possible unconventional superconductivity in NaYbSe$_2$, which belongs to a large and ideal family of triangular lattice spin liquid we revealed recently and is evidenced to possess a QSL ground state. The charge gap of NaYbSe$_2$ is gradually reduced by applying pressures, and at ~20 GPa the crystal jumps into a superconducting (SC) phase with Tc ~ 5.8 K even before the insulating gap is completely closed. The metallization trend remains in the repeated high-pressure experiments but the sign of superconductivity is not well repeated. No symmetry breaking accompanies the SC transition, as indicated by X-ray diffraction and low-temperature Raman experiments under high pressures.

[*] The authors contributed to the work equally.
[‡] yuxh@iphy.ac.cn
[§] qmzhang@ruc.edu.cn

This intrinsically connects QSL and SC phases, and suggests an unconventional superconductivity developed from QSL. We further observed the magnetic-field-tuned superconductor-insulator transition which is analogous to that found in the underdoped cuprate superconductor $La_{2-x}Sr_xCuO_4$. The study is expected to inspire interest in exploring new types of superconductors and sheds light into the intriguing physics from a spin liquid/Mott insulator to a superconductor.

# I. Introduction

The resonating valence bond (RVB) model on the triangular spin lattice was constructed by P. W. Anderson more than four decades ago, and employed to describe a novel spin disordered state QSL, a superposition state of all the possible configurations of valence-bond spin singlets [1]. QSL essentially exhibits strong spin entanglement and hosts a lot of exotic phases/excitations, and can be realized in quantum spin-frustrated systems. Anderson revived the concept of QSL soon after the discovery of cuprate superconductors, and pointed out the ultimate connection between QSL and unconventional superconductivity in cuprates [2]. The parent compounds of cuprate superconductors are antiferromagnetic Mott insulators, and it is believed that the carrier doping pushes them into the QSL phase which eventually drives high-temperature superconductivity. The enormous and unusual electronic properties in the underdoped or pseudogap region can be well described in the framework of QSL [3]. And spin singlets in QSL naturally evolve to Cooper pairs with no additional glue required. Clearly the scenario brings a revolutionary unconventional SC picture beyond BCS mechanism. The experimental verification of the scenario has become one of the central issues in condensed matter physics.

A direct and determinative way to verify the great scenario, is to realize unconventional superconductivity by doping a QSL/Mott insulator. Actually it is really hard to find an appropriate and ideal QSL/Mott compound or material system for a clean and effective carrier doping. Nevertheless, some efforts have been made over the years. The organic compound $\kappa$-(BEDT-TTF)$_2$Cu$_2$(CN)$_3$ ($\kappa$-ET) with an approximate triangular lattice of spin-1/2 Cu$^{2+}$ ions (monoclinic space group P2$_1$/c), was considered to be a QSL candidate since thermodynamics results suggested no magnetic ordering down to 50 mK despite its exchange coupling J ~ 200 K. A very small pressure (~0.35 GPa) can drive a metal-insulator transition at 13 K and a successive SC transition at ~3.9 K [4-6]. It suggests that the compound is a "weak" Mott insulator with a tiny charge gap of ~ 15 meV and the excitations/interactions in spin and charge channels may be easily mixed together. In this sense it may not represent a good example for the doping study. The famous kagome compound ZnCu$_3$(OH)$_6$Cl$_2$ was also employed for doping experiments. But it seems an effective doping was not archived and no metallic phase was observed with Li$^+$ doping, perhaps due to the large charge gap [7]. The lesson from the two compounds leads to the simple but essential conclusion that an effective doping requires a starting QSL/Mott compound with an intermediate charge gap.

Besides 3d-based spin frustrated systems, many rare-earth based QSL compounds have received increasing attention in the last few years. The inorganic YbMgGaO$_4$ discovered in 2015 shows the high symmetry of R-3m [8]. Yb$^{3+}$ ions with an effective S=1/2 guaranteed by Kramers doublets in crystalline electric-field environments, form strict and flat triangular layers which are well separated by GaO/MgO octahedra. The easy availability of high-quality YbMgGaO$_4$ crystals has stimulated extensive experimental investigations and hence theoretical studies. Most experiments including neutron, muSR, thermodynamics etc, point to the conclusion that YbMgGaO$_4$ has the spin ground state of QSL [8-12]. At the same time, the issue on the Ga/Mg disorder was discussed experimentally and theoretically. Just very recently, we revealed a large family of triangular QSL candidates ARECh$_2$ (A=monovalent ion, RE=rare earth, Ch=O, S, Se, Te) named as rare-earth chalcogenides [13]. Exactly analogous to YbMgGaO$_4$, most of the family members possess the high symmetry of R-3m and Yb$^{3+}$ ions form a perfect two-dimensional triangular spin lattice through the network of the tilted and edge-shared YbCh$_6$ octahedra. The susceptibility, specific heat and neutron measurements down to several tens of milikelvins, consistently indicate a QSL ground state at least in NaYbCh$_2$ (Ch=O, S, Se) [14-27]. Most excitingly, the rich diversity of the family members brings a huge selectivity and allows to look for the ideal candidates for doping experiments at a material-engineering level.

The first candidate coming into focus is NaYbSe$_2$, which has been characterized as a QSL compound and most importantly has an appropriate charge gap of ~1.9 eV. We performed the doping studies using high pressure which is recognized as the cleanest method for doping. We checked the structural evolution of NaYbSe$_2$ under high pressures. X-ray diffraction and Raman scattering suggest a subtle and collapse-like structural change with no symmetry change around 12 GPa and no further structural change up to 30 GPa. The resistance is gradually reduced by applying pressures and eventually the crystal enters into a SC phase with Tc ~ 5.8 K at ~20 GPa. By fixing the pressure at ~ 20

GPa, we employed Raman scattering to find no symmetry change across Tc. The metallization trend is confirmed by further high-pressure measurements but oddly the sign of superconductivity is not well repeated. The smooth evolution from a QSL/Mott insulator to SC phase allows to extract a unique phase diagram, with which we are able to discuss some fundamental issues of the transition by making comparison with the organic and cuprate systems.

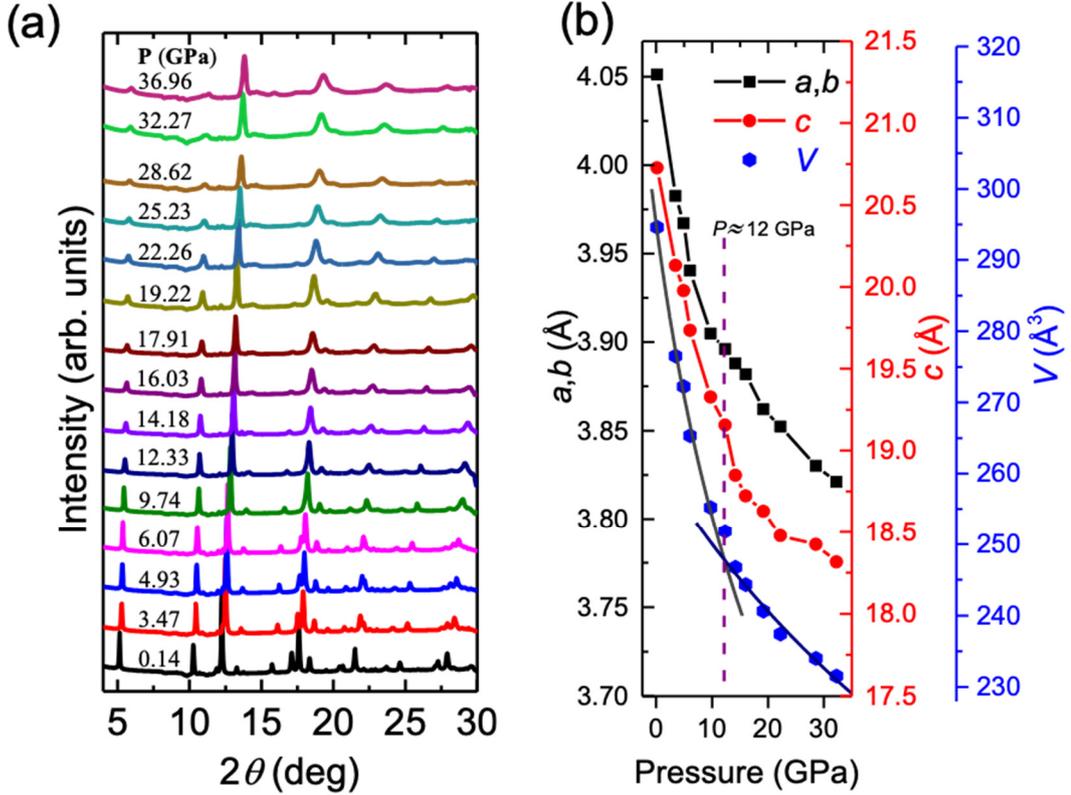

**Fig. 1** (a) High-pressure x-ray patterns for NaYbSe$_2$ crystals. (b) Below 32 GPa, the lattice parameters were refined with *R-3m* (No. 166) symmetry with lattice constants *a*, *b* and *c*. The pressure-dependent cell volume *V* of NaYbSe$_2$ sample was fitted by the third-order Birch–Murnaghan equation: $P(V) = \frac{3B_0}{2}\left[\left(\frac{V_0}{V}\right)^{\frac{7}{3}} - \left(\frac{V_0}{V}\right)^{\frac{5}{3}}\right]\left\{1 + \frac{3}{4}(B_0' - 4)\left[\left(\frac{V_0}{V}\right)^{\frac{2}{3}} - 1\right]\right\}$, where $B_0$ and $B_0'$ (with $B_0'$ fixed to 4) are the isothermal bulk modulus at zero pressure and the pressure derivative of $B_0$ evaluated at zero pressure, respectively; *V* and $V_0$ are the high-pressure volume and zero-pressure volume of the material, respectively. The fitting was conducted below and up 12 GPa, respectively. The dotted line is a guideline for 12 GPa.

## II. Samples and experiments

NaYbSe$_2$ crystals were grown by flux method and self-flux method, and the details can be found elsewhere [13,26].

The high pressure electrical resistance and magnetoresistance measurements were performed by a four-probe method in a diamond anvil cell (DAC). The DAC made of BeCu alloy with two opposing anvils was used to generated high pressure. Diamond anvils with 300 μm culets were used for the measurements. In the experiments, a single crystal sample chip of 50–100 μm was loaded into the sample hole in a gasket pre-indented to ~25–40 μm in thickness, and a few ruby balls were loaded to serve as internal pressure standard. The corresponding gasket was rhenium. KBr was used as pressure-transmitting medium. The ruby fluorescence method was used to determine the pressure [28]. The electrical resistance and magnetoresistance measurements were performed in a physical property measurement system (PPMS).

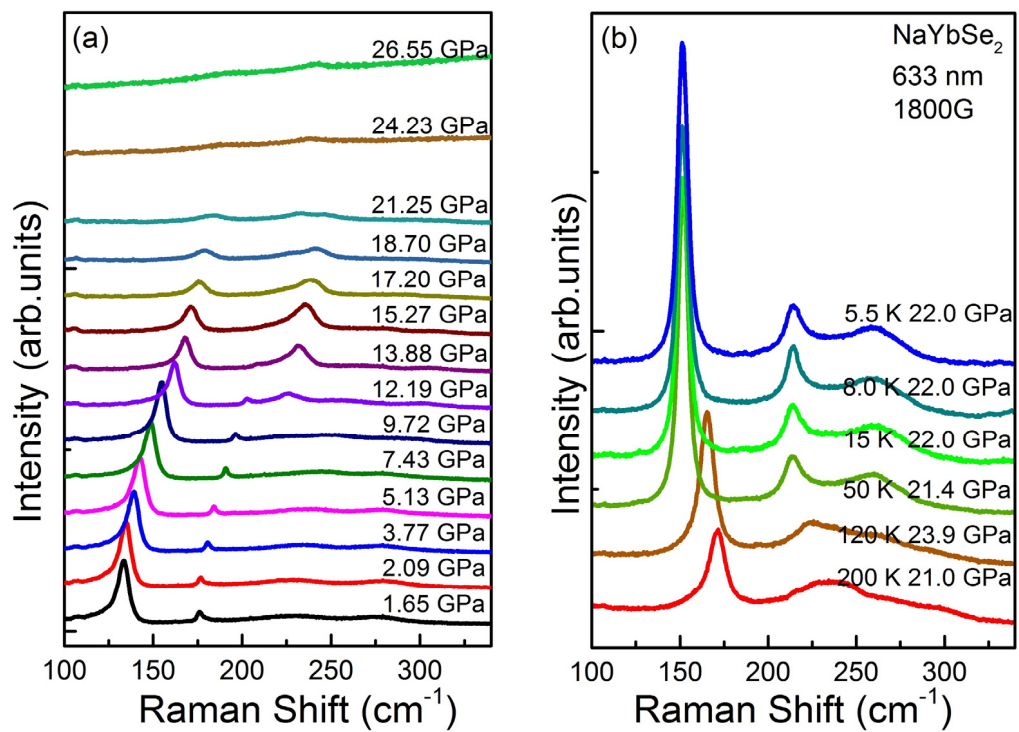

Fig. 2 (a) High-pressure Raman spectra for NaYbSe2 crystals. (b) Temperature-dependent Raman spectra with the pressure fixed at ~21 GPa.

The high-pressure X-ray powder diffraction (XRD) measurements were performed at beamline 4W2 at Beijing Synchrotron Radiation Facility. A monochromatic X-ray beam with a wavelength of 0.6199 Å was used for the measurements. Low birefringence diamonds with culets of 300 μm were used for the experiments. Silicone oil was used as pressure-transmitting medium.

The high-pressure and temperature-dependent Raman measurements presented in this study were performed in gasketed symmetric diamond anvil cells DACs. And the KBr was used as pressure-transmitting medium in Raman experiments. The Raman spectra were collected using a HR800 spectrometer (Jobin Yvon) equipped with a liquid-nitrogen-cooled charge-coupled device (CCD) and volume Bragg gratings, for which micro-Raman backscattering configuration was adopted. A 633 nm laser was used, with a spot size of~5μm focused on the sample surface. The laser power was maintained at approximately 1.4 mW to avoid overheating during measurements.

### III. Results and discussions

Fig. 1 shows the structural evolution with pressure up to 37 GPa. In general, X-ray patterns exhibit a relatively smooth evolution with pressure (Fig. 1a). A slight kink in the parameter *c* and the volume V of unit cell, appears around 12 GPa (Fig. 1b). It is a subtle structural change which can be attributed to a collapse along c-axis without symmetry change since the parameters of *a* and *b* pass through the pressure point smoothly (Fig. 2c). This is also consistent with Raman experiments under high pressure (Fig. 2). It should be pointed out that there is also no sign of structural change at ~ 20 GPa where superconductivity is found (See below).

Raman spectra under high pressure are shown in Fig. 2. There are two Raman allowed modes at ~130 and 175 $cm^{-1}$ and at low pressures. They smoothly move to higher frequencies with pressure. There appears a stronger peak around 220 $cm^{-1}$ above 12 GPa. It is a crystalline electric-field excitation rather than a new phonon mode, as identified elsewhere [26]. It means that despite a subtle structural collapse at ~12GPa, the crystal symmetry remains unchanged up to 37 GPa, as suggested by X-ray results under high pressure discussed above. Particularly at ~20 GPa which is the critical pressure for superconducting transition, Raman spectra, just like X-ray patterns, show no anomaly. If we fix the pressure at ~20 GPa, the temperature-dependent Raman spectra clearly tell us there is no structural change at least in the range of 5.5 K (below Tc) to 210 K (Fig. 2b).

The metallization under pressure is directly illustrated in Fig. 3a, in which the crystal color becomes more and more dark with pressure, indicating the gradual increase of metallicity. This is also clearly evidenced in the R-T curves in Fig. 3b and 3c. As mentioned above, the pristine NaYbSe$_2$ has a charge gap of ~1.9 eV, which is why the crystal looks brown. At pressures lower than 18.5 GPa, the crystal behaves like a good insulator with an exponentially divergent resistance at low temperatures (Fig. 3b). The resistance falls into the measurable range above 18.8 GPa (Fig. 3c). Interestingly, the resistance first shows a temperature dependence with a positive slope featureing a Fermi liquid-like character, then turns to an insulating state with a negative slope below a characteristic temperature. The temperature can be extracted with dR/dT=0 and essentially characterizes the crossover from a high-T/high-P Fermi liquid to a low-T/low-P spin liquid/Mott insulator. We will establish a temperature-pressure phase diagram later and have more discussions there.

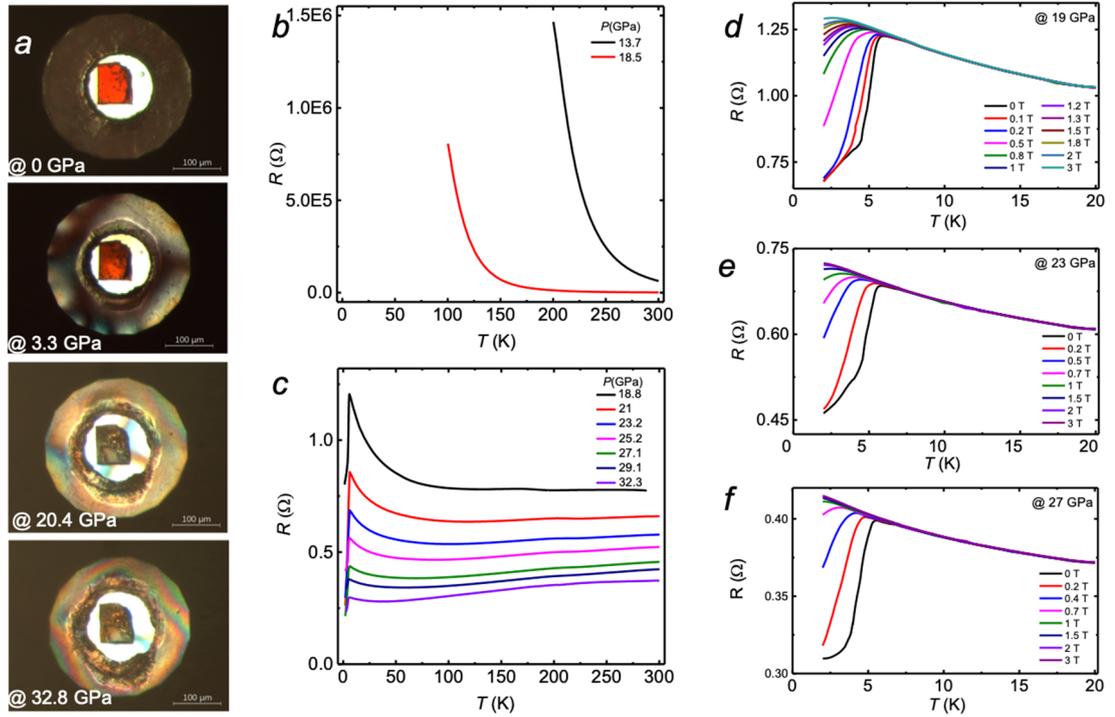

**Fig. 3** (a) Color evolution of NaYbSe$_2$ crystals under various pressures. (b) & (c) Temperature-dependent resistance under pressure. (d), (e) & (f) Evolution of SC transition under magnetic fields with the pressure fixed at 19, 23 and 27 GPa, respectively.

The most exciting observation is the emergence of superconductivity. At a starting pressure of ~18.8 GPa, the crystal shows a sudden drop in resistance at ~5.8 K (Fig. 3c), which is confirmed to be a SC transition by the Meissner-effect measurements under magnetic fields though a zero resistance is not achieved (Fig.

3d, 3e, 3f). The pressure-induced superconductivity is quite unusual in several aspects. First, the SC phase is in close proximity to the insulating one and it seems the insulating phase jumps into the SC phase without an intermediate metallic state. Second, the transition temperature Tc remains almost unchanged with further increasing pressures. Finally, the upper critical field is quite low (<3T) and also almost unchanged with pressure (Fig. 3d, 3e, 3f). We will work out a phase diagram later, with which we can have more words on the underlying physics related to the superconductivity.

We have tried more high-pressure measurements to repeat the above observations. The reduction of resistance with pressure is clear for each measurement (Fig. 4a and 4b). On the other hand, we have to honestly point out that the SC transition is not well repeated. At the moment we have no good angle to understand the negative results. If the superconductivity could be extrinsic and come from the compounds other than $NaYbSe_2$, the elemental Yb out of the elemental Na, Yb and Se and their combinations, could be the most plausible candidate. However, superconductivity in elemental Yb occurs at ~4.6 K under a pressure higher than 179 GPa [29]. It simply rules out the possibility. Clearly this is an open issue and more extended studies are highly required.

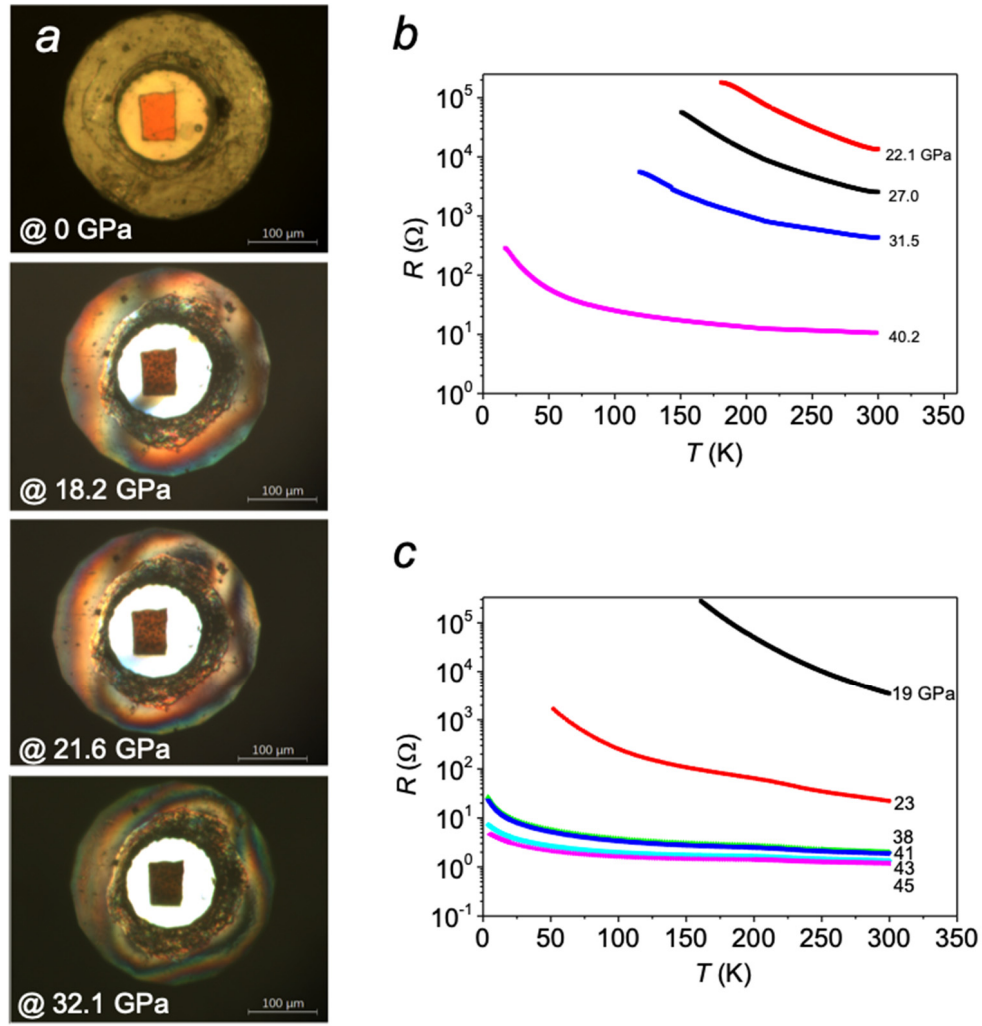

**Fig. 4** (a) Color evolution of NaYbSe$_2$ with pressure. (b) & (c) Two successive confirmation high-pressure measurements. Metallization trend or resistance reduction with increasing pressure is clearly seen again while the SC transition is not well repeated in the measurements.

Despite the negative results, the transition temperature Tc and the characteristic temperatures extracted from the resistance curves at dR/dT=0, allow us to establish a temperature-pressure phase diagram (Fig. 5). The characteristic temperatures form a boundary between Ferimi liquid and spin liquid/Mott insulator. Generally a very small doping will drive a spinon system into a SC state in term of the picture of quantum spin liquid [3]. But strong thermal fluctuations at higher temperatures prevent the SC condensation and keep the system in the metallic phase. In this sense, the Fermi-liquid phase in the phase diagram may be inaccurate since it looks more like the pseudogap region in the cuprate superconductors. It would be expected to dig out exotic features in the metallic state, considering its elementary excitations could be fundamentally distinct from normal electrons in Fermi liquid.

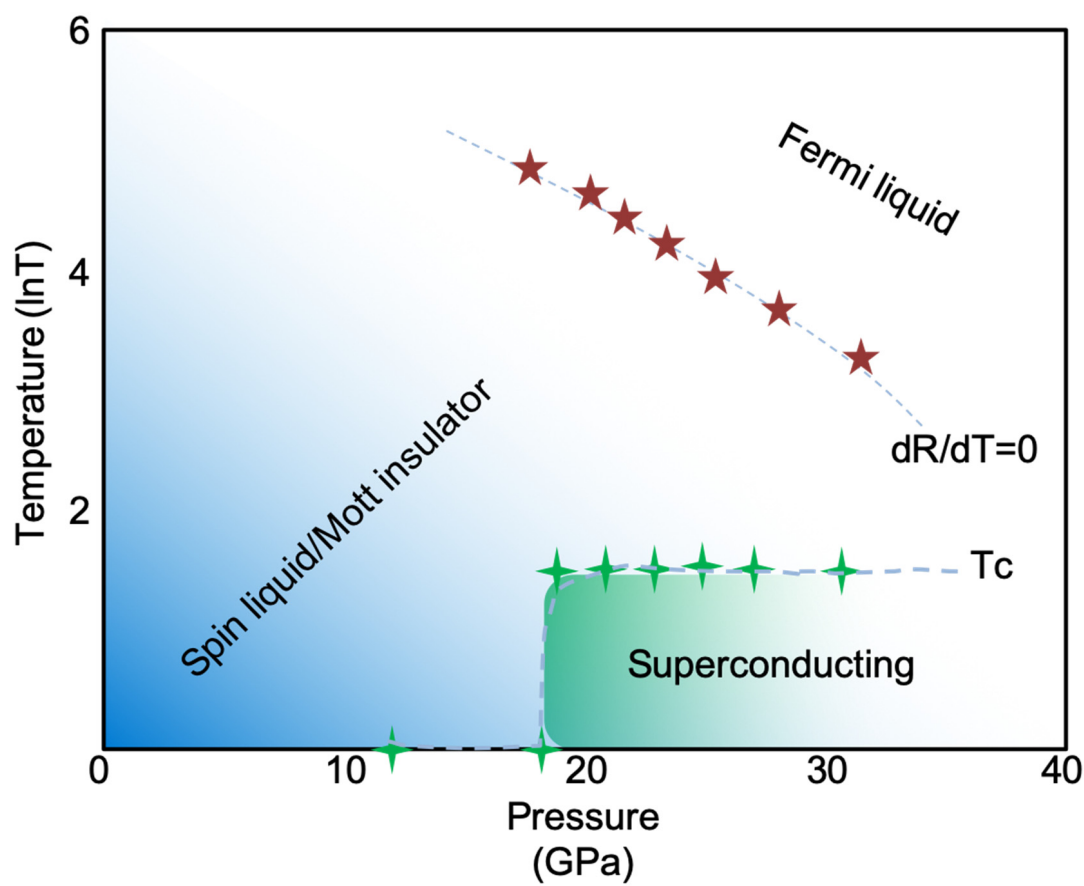

**Fig. 5** Phase diagram of pressure vs. temperature. The pentastars are extracted from Fig. 3c with dR/dT=0 and the cross-stars are transition temperatures obtained also from Fig. 3c.

Tc forms another boundary to separate the SC phase from the spin liquid/Mott insulator. First of all, we need to emphasize that no spatial symmetry breaking occurs across the boundary. This is supported by X-ray and Raman measurements under high pressures discussed above. In other words, the SC phase is developed from spin liquid/Mott insulating phase, and both phases are intrinsically connected. This is hinted by the experimental facts. As we mentioned above, a very low doping can drive a spin liquid to a superconductor. Actually in the theoretical limit, even one single mobile carrier doping into a spin liquid can give rise to a SC state. Experimentally this is reflected on the close proximity to the insulating state. And the low upper critical fields correspond to a low superfluid density which is also consistent with the above picture.

More interesting features come out when we compare the present system to organic and cuprate superconductors. The organic $\kappa$-ET$_2$Cu$_2$CN$_3$ is also considered to possess a parent phase of spin liquid/Mott insulator, and can realize a SC transition under pressure [4-6]. One key difference between $\kappa$-ET$_2$Cu$_2$CN$_3$ and NaYbSe$_2$ is the intermediate metallic state. Before going into the SC state, the organic compound always undergoes an intermediate metallic state along the temperature axis when fixing pressure, while in our case there is no such a metallic phase along either temperature or pressure axis. In this sense, the case of NaYbSe$_2$ looks more compatible with the original picture. On the other hand, in what conditions the metallic phase appears and how its excitations behave are interesting issues. The magnetic field dependence of the SC transition, is another surprising observation. The magnetic field gradually suppresses the SC phase and eventually drives it into an insulating phase rather than a metallic phase (Fig. 3d, 3e, 3f). The magnetic field tuned superconductor-insulator transition was exactly observed in the underdoped cuprate superconductor La$_{2-x}$CuO$_4$ [30]. The interesting similarity allows to look into the SC mechanism in cuprates in a new angle.

Generally a SC transition can be described as gauge symmetry breaking in term of Anderson-Higgs mechanism. Here the parent phase is a spin liquid with additional emergent gauge field rather than a conventional metal. The question is whether or not the emergent gauge field is just simply confined in the transition from spin liquid/Mott insulator to superconductor, or we still have chance to look into the emergent gauge structure in some way. This is a fundamental issue which may find some hints on the present unique platform.

### IV.    Summary

We made high-pressure study on the spin liquid candidate NaYbSe$_2$. Pressure induced metallization is clearly observed and the indication of possible unconventional superconductivity is also presented though the SC transition is not well repeated in the successive confirmation measurements. The SC transition is accompanied by no spatial symmetry breaking and the SC phase is intrinsically evolved from the unusual parent phase of spin liquid/Mott insulator. Many unusual features like the close proximity to an insulating phase, almost pressure independent transition temperature and low upper critical field etc, are revealed. As we mentioned before, rare-earth chalcogenides form a large and ideal family of triangular lattice spin liquid. The present study opens a window to look into the intriguing physics from spin liquid/Mott insulator to superconductor, and is just a first step towards potential extensive experimental and theoretical studies on some fundamental issues in condensed matter physics.

## Acknowledgements

We would like to point out that we become aware of the nice high-pressure experiments posted on arXiv very recently during the preparation of the present manuscript [31]. This work was supported by the Ministry of Science and Technology of China (2017YFA0302904 & 2016YFA0300500) and the NSF of China (11774419, U1932215 & 11974244).